\documentclass[epj,nopacs]{svjour}
\usepackage{graphicx}
\usepackage{caption}
\usepackage{color}
\usepackage{amsmath}
\usepackage{amssymb}
\usepackage{amsfonts}
\usepackage{bm}
\usepackage{hyperref}
\usepackage{url}
\usepackage{subcaption}
\usepackage{mathtools}
\usepackage{cuted}
\usepackage[english]{babel}
\usepackage[left=2cm,right=2cm,top=2cm,bottom=2cm]{geometry}
\newcommand{\bfr}{{\bm {r}}}
\newcommand{\bfp}{{\bm {p}}}

\newcommand{\bfb}{{\bm {b}}}
\newcommand{\bfe}{{\bm {\eta}}}
\newcommand{\bfq}{{\bm {q}}}

\newcommand{\hp}{\hat{\bfp}}

\newcommand{\hq}{\hat{\bfq}}

\newcommand{\la}{\langle}
\newcommand{\ra}{\rangle}

\newcommand{\eps}{\ensuremath{\varepsilon}}
\newcommand{\kap}{\ensuremath{\kappa}}

\newcommand{\vphi}{\ensuremath{\varphi}}

\renewcommand{\vec}[1]{{\mbox{\boldmath$#1$}}}
\begin{document}
\thispagestyle{empty}
\title{Doubly differential cross sections for ionization of lithium atom by protons and O$^{8+}$ ions}
\author{Andrey~I.~Bondarev\inst{1,2,3} \and Yury~S.~Kozhedub\inst{2,3} \and Ilya~I.~Tupitsyn\inst{1,2} \and Vladimir~M.~Shabaev\inst{2} \and G\"unter~Plunien\inst{4}}
\institute{
Center for Advanced Studies, Peter the Great St. Petersburg Polytechnic University, 
Polytechnicheskaya 29, 195251 St. Petersburg, Russia 
\and
Department of Physics, St. Petersburg State University, 
Universitetskaya 7/9, 199034 St. Petersburg, Russia 
\and
NRC ``Kurchatov Institute'',
Academician Kurchatov 1, 123182 Moscow, Russia
\and 
Institut f\"ur Theoretische Physik, Technische Universit\"at Dresden,
Mommsenstrasse 13, D-01062 Dresden, Germany
}
\mail{bondarev@spbstu.ru}
\abstract{
We consider single ionization of lithium atom in collisions with $p$ and O$^{8+}$ projectiles. Doubly differential cross sections for ionization are calculated within a relativistic non-perturbative approach. Comparisons with the recent measurements and theoretical predictions are made.
}
%
\authorrunning{Bondarev {\it et al.}}
\titlerunning{DDCS for ionization of Li by $p$ and O$^{8+}$ projectiles}
\maketitle
\section{Introduction}
%
During last decades the cold target recoil ion momentum spectroscopy (COLTRIMS)~\cite{doe_00,ull_03} has been widely applied to study break-up processes of simple atomic and molecular systems~\cite{ull_04}. With this technique, also known as a reaction microscope (ReMi), the momenta of ejected electrons and recoiling target fragments are measured directly, while the final projectile momentum is obtained from the conservation laws. In this way, kinematically complete experiments, in which fully differential cross sections can be addressed, are feasible. The range of target species for the COLTRIMS experiments is restricted due to the employed supersonic gas-jet technique. The most part of the studies was done with helium and molecular hydrogen targets, for which the best momentum resolution can be reached. These limitations were overcome by a combination of a magneto-optical trap (MOT) for target cooling and a ReMi. The constructed MOTReMi setup is described in detail in Ref.~\cite{hub_15}. Furthermore, a successful implementation of an all-optical trap (AOT) in the MOTReMi experiment has been recently reported~\cite{sha_18}. Unlike conventional MOTs, the AOT does not require magnetic field gradients in the trapping region. This feature greatly facilitates the joint operation of the trap and ReMi.
Within the MOTReMi technique, the differential cross sections for single ionization of lithium by $6$~MeV protons and $1.5$~MeV/u O$^{8+}$ were measured~\cite{fis_12,hub_13,LaF_13}. Since that time several theoretical calculations were performed in order to explain the experimental data. The continuum-distorted-wave eikonal-initial-state (CDW-EIS) method was applied in Ref.~\cite{gul_14} to calculate differential cross sections in these collisions. The role of multielectron processes was investigated in Refs.~\cite{kir_14,spi_16} also within the CDW-EIS framework. In the latter paper, the CDW-EIS approach was used in combination with the two-center basis generator method~\cite{zap_05}. The comparison between advanced perturbative theories, three-body continuum-distorted-wave (3DW) and three-body continuum-distorted-wave eikonal-initial-state (3DW-EIS), was reported in Ref.~\cite{gha_16}. Large discrepancies between the results of the CDW-EIS and 3DW-EIS approaches at large momentum transfers were recently found in Ref.~\cite{gno_17}. Non-perturbative calculations of the differential cross sections were carried out by the time-dependent close coupling (TDCC) and coupled-pseudostate (CP)  approaches in Ref.~\cite{cia_13} and Ref.~\cite{wal_14}, correspondingly. 
In this contribution, we apply the relativistic non-perturbative approach of Ref.~\cite{bon_17} to calculate doubly differential cross sections (DDCS) for single ionization of lithium atom by protons and bare oxygen nuclei. 
The article is organized as follows. In Sec.~\ref{s:theory}, we summarize the key points of the method developed in Ref.~\cite{bon_17} and emphasize features of the lithium target description. In Sec.~\ref{s:results}, the results of calculations are presented and discussed. The conclusions are drawn in Sec.~\ref{s:conclusion}. Atomic units (a.u.) $\hbar = e = m_{e} = 1$ are used throughout the paper unless otherwise stated. 
%
\section{Theory} \label{s:theory}
%
General formulation of the approach has been given in Ref.~\cite{bon_17}. In Ref.~\cite{bon_18}, it was applied to the essentially three-body collision of antiproton with atomic hydrogen. Here, we use the same approach to study single ionization of Li by protons and O$^{8+}$ ions. We describe a lithium target atom in the one-active-electron approximation, where the only $L$-shell electron is active, while two $K$-shell electrons belong to the frozen core. The interaction potential $V_{\rm T}$ between the active electron and the core is calculated using the density-functional approximation with self-interaction correction~\cite{per_81}. The effectively three-body collision of a projectile with a target composed of the frozen core and the active electron is considered in the impact parameter representation, where the projectile moves along a straight-line trajectory $\vec{R} = \vec{b}+\vec{v}t$ with the constant velocity $\vec{v}$ and at the impact parameter $\vec{b} = (b,\phi_b)$, so that $\vec{b}\cdot\vec{v} = 0$.
The wave function of the active electron $\Psi$ obeys the time-dependent Dirac equation
\begin{equation} \label{eq:dirac}
i\frac{\partial \Psi(\vec{r},t,\vec{R})}{\partial t} = \Bigl[H_{\rm T}+V_{\rm P}(t)\Bigr]\Psi(\vec{r},t,\vec{R}), 
\end{equation}
where the stationary target Hamiltonian $H_{\rm T}$ comprises the kinetic energy and the interaction between the active electron and the target core,
\begin{equation} \label{eq:t_hamiltonian}
H_{\rm T} = c(\vec{\alpha} \cdot \vec{p})+(\beta-1) c^2+V_{\rm T}
\end{equation}
with $\vec{\alpha}$ and $\beta$ being the Dirac matrices, $c$ is the speed of light.
The interaction between the active electron and the projectile is
\begin{equation} \label{eq:p_potential}
V_{\rm P} = -\frac{Z_{\rm P}}{|\vec{r}-\vec{R}|}, 
\end{equation}
where $Z_{\rm P}$ is the charge of the projectile.
In the impact-parameter approximation, the interaction between the projectile and the target core, so called nucleus-nucleus (NN) interaction can not change the predetermined trajectory. Therefore, for cross sections not differential in the scattered projectile variables, it can be omitted in Eq.~\eqref{eq:dirac}. Since the NN interaction does not depend on the electronic coordinates, it can be removed  from (or added to) Eq.~\eqref{eq:dirac} by a corresponding choice of the wave function phase. This phase, however, should be taken into account in the calculation of the cross sections differential in the scattered projectile variables. Though, this transformation should be done with caution~\cite{wal_12}.
Expanding the time-dependent wave function $\Psi$ over a stationary finite basis set $\{\vphi_a\}$,
\begin{equation} \label{eq:basis_exp}
\Psi(\vec{r},t, \vec{R}) =  \sum_{a}C_{a}(t,\vec{b}) e^{-i\eps_{a}t}\vphi_{a}(\vec{r}),
\end{equation}
and substituting Eq.~\eqref{eq:basis_exp} into Eq.~\eqref{eq:dirac}, we derive the set of coupled-channel  equations for the time-dependent expansion coefficients $C_{a}(t,\vec{b})$,
\begin{equation}  \label{eq:cc}
i\frac{dC_a(t,\vec{b})}{dt} = \sum_{b} C_{b}(t,\vec{b}) e^{i(\eps_a-\eps_b)t }\la\vphi_a|V_{\rm{P}}|\vphi_b\ra.
\end{equation} 
The basis functions $\vphi_a$ form an orthonormal set. They are obtained by diagonalization of the target Hamiltonian $H_{\rm T}$ on {\it B}-splines~\cite{joh_88,bon_15},
\begin{equation} \label{eq:basis_const}
\la \vphi_a | H_{\rm T} | \vphi_b \ra = \eps_a\delta_{ab}, \qquad \la \vphi_a | \vphi_b \ra = \delta_{ab}.
\end{equation}
The system of the coupled channel equations~\eqref{eq:cc} is solved subject to the initial conditions
\begin{equation} \label{eq:init_cond}
C_a(t \to-\infty,\vec{b}) = \delta_{ai}.
\end{equation}
It is worth noting that the atomic-like basis set centered at the target is not suitable for the explicit description of charge exchange processes. If these processes are significant, their contribution is also included into the ionization, which is, in fact, the electron loss from the target. Two-center basis sets should be used in order to take into account charge transfer processes explicitly~\cite{tup_12,wal_15,abd_16a}.

Having the set of expansion coefficients $\{C_{a}(t,\vec{b})\}$ at asymptotic time $t \to \infty$, we extract information about active electron transitions during the collision. The transition amplitude to the state with a given energy $\eps$, asymptotic momentum direction $\hp = \vec{p}/p =(\theta_e, \phi_e)$, helicity~$\mu_s$, and incoming spherical waves boundary conditions reads as~\cite{bon_17}  
\begin{equation} 
T^{\mu_s}(\eps, \theta_e, \phi_e, \vec{b}) = \la \Psi^{(-)}_{\eps \hp \mu_s}\, e^{-i\eps t} | \Psi(t) \ra, \quad t \to \infty.
\end{equation} 
Using a Fourier transform, we can express the transition amplitude in the representation of the transverse component $\vec{\eta} = (\eta,\phi_\eta)$ of the momentum transfer $\vec{q} = \vec{k_i}-\vec{k_f}$ with $\vec{k_i}$ ($\vec{k_f}$) being the initial (final) projectile momentum~\cite{eic_95}, 
\begin{equation} \label{eq:FT_gen}
T^{\mu_s}(\eps,\theta_e,\phi_e,\vec{\eta}) = \frac{1}{2\pi}\int \! d\vec{b}\, e^{i\bfe\cdot\bfb}\, e^{i\delta(b)}\,  T^{\mu_s}(\eps,\theta_e,\phi_e,\vec{b}),
\end{equation}
where $\delta(b)$ is the additional phase due to the NN interaction omitted in Eq.~\eqref{eq:dirac}.
Different types of the semi-empirical NN potentials were examined in Ref.~\cite{gul_14}. It was found that at small momentum transfers, the DDCS is sensitive solely to the asymptotic form of the NN potential at large internuclear distances. In this work, we approximate the NN interaction by the simplest expression with the correct large-distance asymptotics,
\begin{equation}  \label{eq:NNeff}
V_{\rm NN}(R) = \frac{Z_{\rm eff}Z_{\rm P}}{R},
\end{equation}
where $Z_{\rm eff} = 1$. The same approximation for the NN interaction has been used in the TDCC calculation~\cite{cia_13}.
Squaring the absolute value of the amplitude~\eqref{eq:FT_gen} and summing over two different helicities $\mu_s$ give us the fully differential ionization probability as a function of the transverse component of the momentum transfer $\vec{\eta}$, the electron ejection energy $\eps$, and the electron ejection angles $\theta_e$ and $\phi_e$,
\begin{equation}  \label{eq:d3P_eta}
\frac{d^3P(\vec{\eta})}{d\eps\, d(\cos\theta_e)\, d\phi_e} = \sum_{\mu_s = \pm\frac{1}{2}}|T^{\mu_s}(\eps, \theta_e, \phi_e, \vec{\eta})|^2.
\end{equation} 
The cross section for the electron being ejected with the energy in the range from $\eps$ to $\eps+d\eps$ into the solid angle $d\Omega_e$, while the projectile is scattered into the solid angle $d\Omega_{\rm P}$ is given by
\begin{equation} \label{eq:fdcs-eta}
\frac{d^3\sigma}{d\eps\, d\Omega_e\, d\Omega_{\rm P}} = k_i k_f \frac{d^3P(\vec{\eta})}{d\eps\, d(\cos\theta_e)\, d\phi_e}.
\end{equation}
This triply differential cross section (TDCS) is different in the laboratory and center of mass reference frames as the projectile momenta $k_i$ and $k_f$ and scattering angle $d\Omega_{\rm{P}}$ also depend on the frame. Here, we focus on the DDCS $\frac{d^2\sigma}{d\eps \, d\eta}$ differential in energy of the ejected electron and transverse component of the projectile momentum transfer, which is obtained by integration of the TDCS,
\begin{equation} \label{eq:d2sigdedeta_def}
\frac{d^2\sigma}{d\eps\, d\eta} = \frac{\eta}{k_i k_f} \int_0^{2\pi} \!\! d\phi_{\rm P} \int \! \frac{d^3\sigma}{d\eps\, d\Omega_e\, d\Omega_{\rm P}} d\Omega_e,
\end{equation}
where $\phi_{\rm P}$ is the azimuthal angle of the scattered projectile. Using Eq.~\eqref{eq:fdcs-eta}, the DDCS can be calculated as
\begin{equation} \label{eq:d2sigdedeta_cal}
\frac{d^2\sigma}{d\eps\, d\eta} = \eta \int_0^{2\pi} \!\! d\phi_\eta\, \int_{-1}^{1} \!\! d(\cos\theta_e) \int_0^{2\pi} \!\! d\phi_e \frac{d^3P(\vec{\eta})}{d\eps\, d(\cos\theta_e)\, d\phi_e}.
\end{equation} 
This cross section does not depend on the reference frame.

The first Born approximation (FBA) serves as a simple perturbation benchmark to evaluate the non-perturbative effects. Furthermore, it helps to verify the convergence of the time-dependent coupled-channel calculation in the basis set size and other parameters. By turning off the couplings with all but the ground state in the right-hand-side of Eq.~\eqref{eq:cc}, the time-dependent calculation can be run in the first Born mode. The comparison of its result with the outcome of the wave treatment FBA provides the convergence estimate. Unlike the case of ionization of a hydrogenlike ion, where the first Born ionization amplitude is known analytically~\cite{mcd_70}, here, this amplitude has to be obtained numerically. Since our approach is relativistic, the ionization amplitude and cross section calculated employing the relativistic wave functions are needed as well. We begin with the well-known expression for the first Born amplitude 
\begin{equation}  \label{eq:f_ion_FBA}
T^{\rm FBA} = \frac{2 Z_{\rm P}}{q^2}\la\psi^{(-)}_{\eps \hp \mu_s}|e^{i\bfq\bfr}|\psi_i\ra,
\end{equation}
where $\psi^{(-)}_{\eps \hp \mu_s}(\vec{r})$ and $\psi_i(\vec{r}$) are the wave functions of the final and initial state, correspondingly. In our case, these functions are represented by bispinors. Expanding the wave function $\psi^{(-)}_{\eps \hp \mu_s}(\vec{r})$ and plane wave $e^{i\bfq\bfr}$ in partial waves, after some algebra we arrive at
\begin{strip}
\begin{multline} \label{eq:pw_exp}
T^{\rm FBA} = \frac{8\pi Z_{\rm P}}{q^2}\sum_{j=1/2}^{\infty}\;\sum_{\mu=-j}^{j}\;\sum_{l=j-1/2}^{j+1/2}\;\sum_{\lambda=|j-j_i|}^{j+j_i}i^{\lambda-l}e^{i\Delta_{jl}}\sqrt{\frac{2\lambda+1}{4\pi}}C^{j\mu}_{l\mu-\mu_s,1/2\mu_s}g^{\lambda \mu_i-\mu}(j\mu;j_i \mu_i) 
Y_{l\mu-\mu_s}(\hp)Y^{*}_{\lambda \mu_i-\mu}(\hq) \\
\times \int_0^\infty dr j_{\lambda}(qr)[G_{\eps\kap}(r)G_{n_i \kap_i}(r)+F_{\eps\kap}(r)F_{n_i \kap_i}(r)].
\end{multline}
\end{strip}
Here $C^{j\mu}_{lm,s\mu_s}$ and $g^{LM}(j_a \mu_a;j_b \mu_b)$ are the Clebsh-Gordan and relativistic Gaunt coefficients~\cite{bon_17}, respectively; $Y_{LM}$ and $j_{\lambda}$ are the spherical harmonics and spherical Bessel functions of the first kind, respectively; $G(r)$ and $F(r)$ are the large and small radial components of the wave functions, respectively; $\Delta_{jl}$ is a phase shift. Thus, treating Eq.~\eqref{eq:pw_exp} with a sufficient number of partial waves $\kappa = \{jl\}$, one can calculate the FBA amplitude with a required accuracy. Due to behavior of Bessel functions of a large order at small arguments, the sum over partial waves in Eq.~\eqref{eq:pw_exp} converges quite fast.
A similar derivation for the non-relativistic case was used in Ref.~\cite{wal_14}, and the corresponding result there was labeled EXB1.
Finally, the FBA TDCS in the center of mass reference frame is given by
\begin{equation}  \label{eq:fdcs_def}
\frac{d^3\sigma}{d\eps\, d\Omega_e\, d\Omega_{\rm P}} = p\frac{k_f}{k_i}\mu^2 \!\!\!\sum_{\mu_s=\pm 1/2}|T^{\rm FBA}|^2,
\end{equation} 
where $\mu$ is the reduced mass of the system. 
The desired DDCS is calculated according to Eq.~\eqref{eq:d2sigdedeta_def}.
\section{Results}  \label{s:results}
The energy of the $2s$ ($2p$) state of lithium calculated by diagonalization of the target Hamiltonian $H_{\rm T}$ employing {\it B} splines equals to $-5.35$ eV ($-3.65$)~eV, which is in fair agreement with experimental values~\cite{NIST}.
In the relativistic treatment, six $2p$ electrons are characterized by quantum numbers \{$j\mu$\}, where $j=1/2,3/2$ and $\mu=-j,...,j$. While the DDCS does not depend on the sign of $\mu$, it depends on the absolute values of both $j$ and $\mu$. In the calculations, we have assumed the equal population of the $2p_{1/2}(1/2)$, $2p_{3/2}(1/2)$, and $2p_{3/2}(3/2)$ states.
Since the experimental data of Refs.~\cite{hub_13,LaF_13} are not on the absolute scale, we have normalized them to our non-pertubative results for $2$-eV ejection from Li($2s$) at transverse momentum transfer $\eta=0.65$~a.u. This normalization procedure was also used in the previous studies~\cite{LaF_13,cia_13,wal_14,gul_14}.
%
\subsection{Proton-impact ionization}  \label{s:proton-imp}
\begin{figure}[ht]
    \centering
    \begin{subfigure}{0.48\textwidth}
        \includegraphics[width=\textwidth]{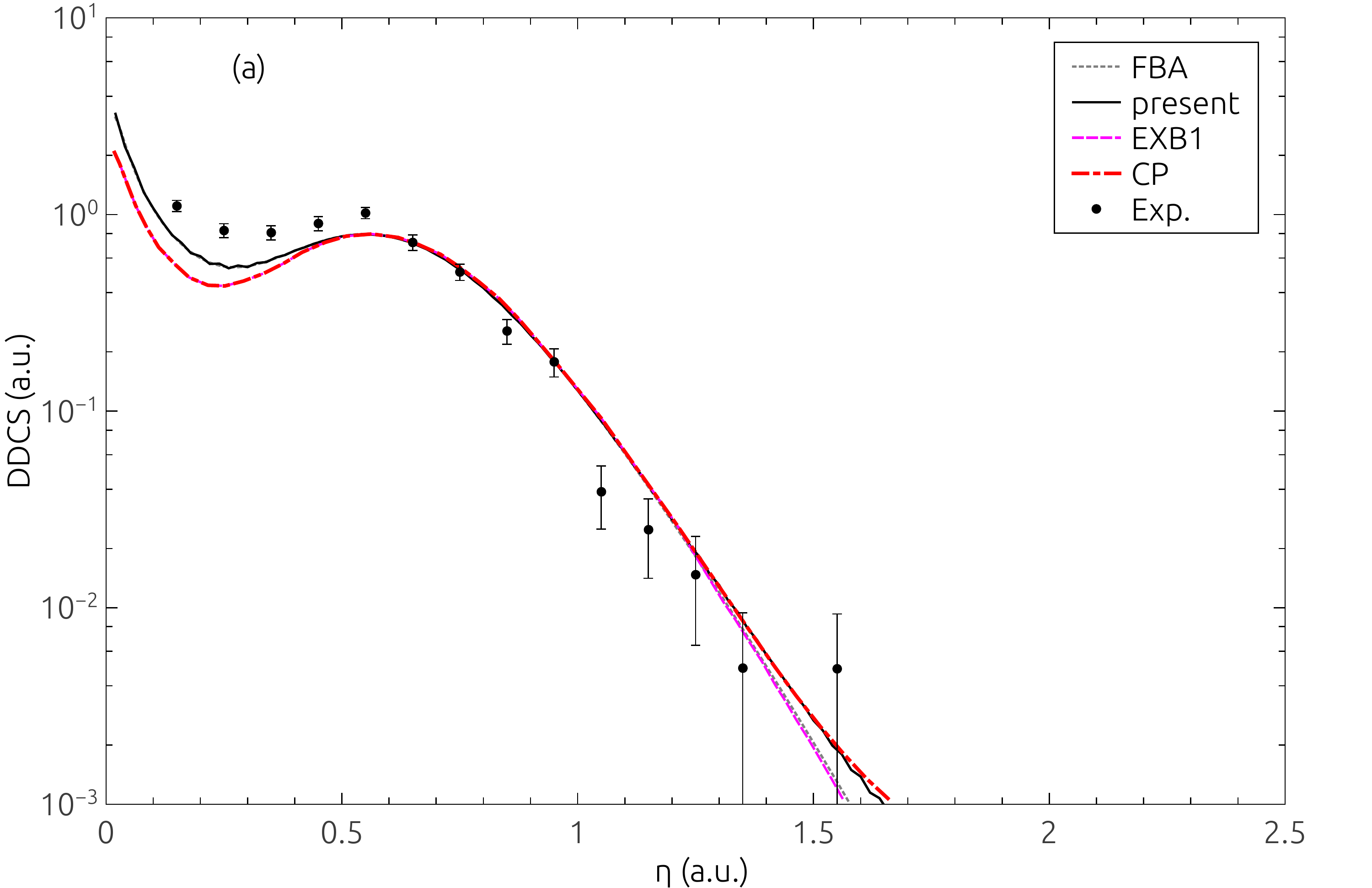}
    \end{subfigure}
\\ 
    \begin{subfigure}{0.48\textwidth}
        \includegraphics[width=\textwidth]{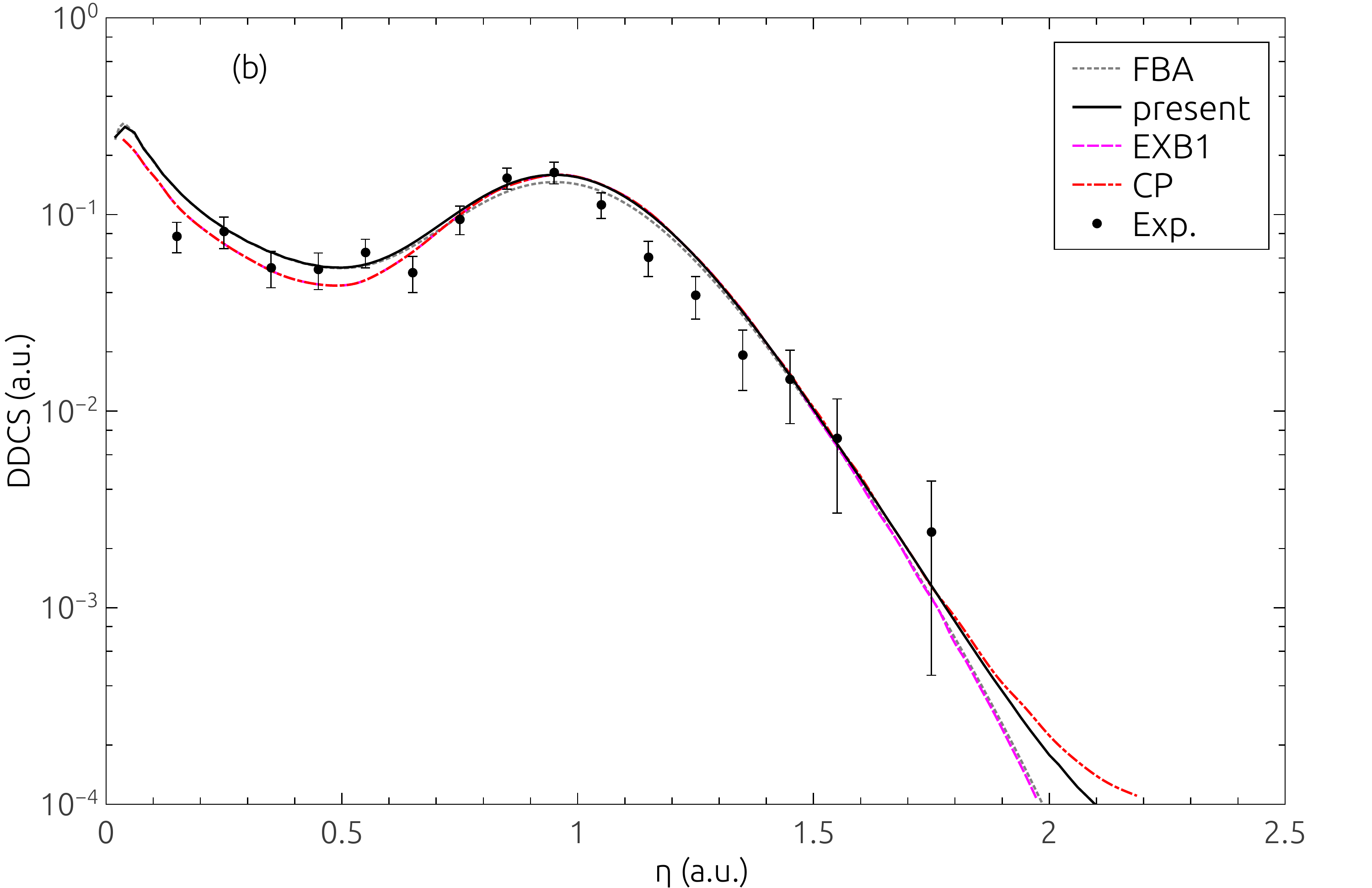}
    \end{subfigure}
\\
    \begin{subfigure}{0.48\textwidth}
        \includegraphics[width=\textwidth]{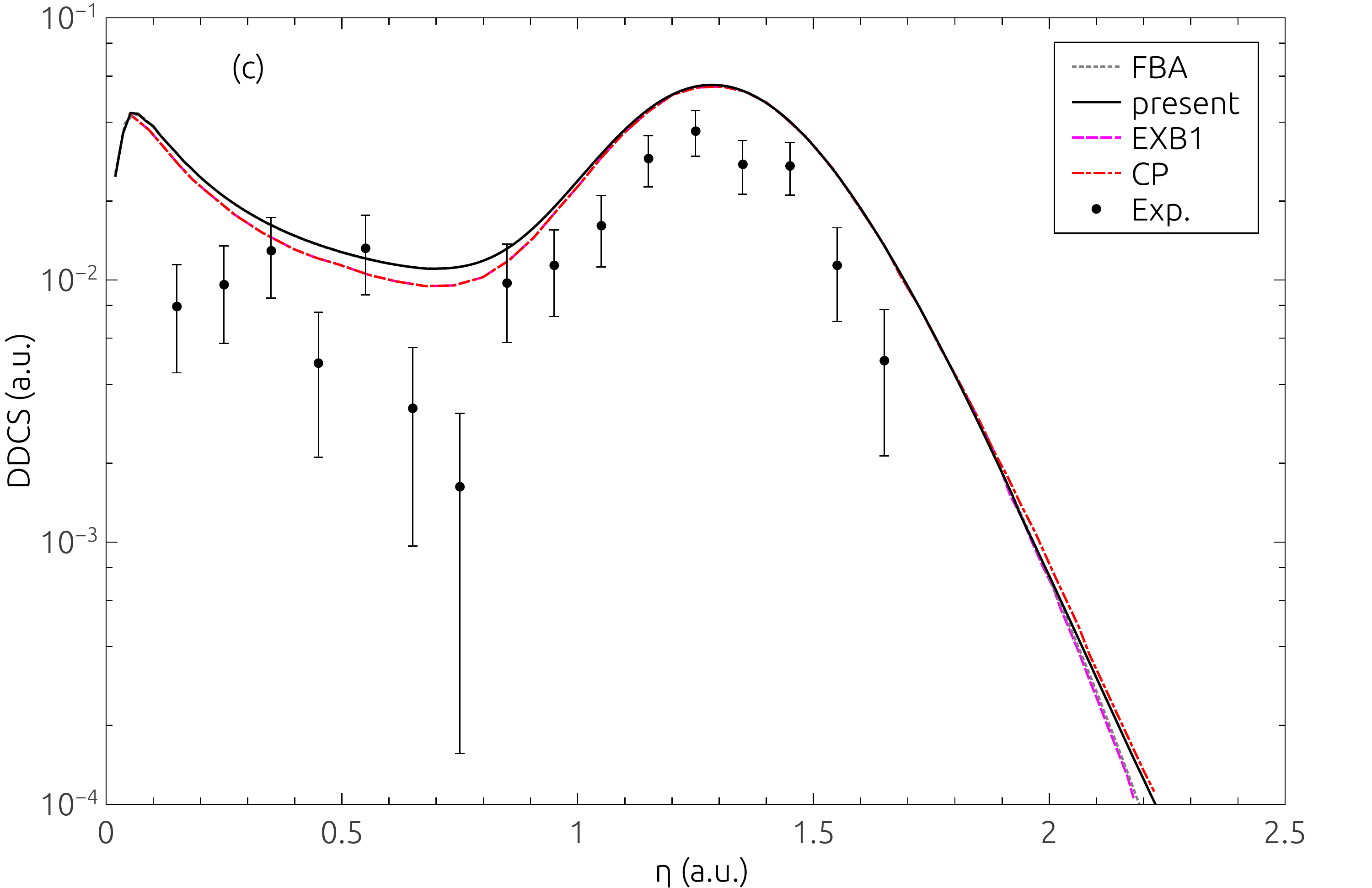}
    \end{subfigure}
    \caption{DDCS $\frac{d^2\sigma}{d\eps\, d\eta}$ as a function of the transverse component of the momentum transfer $\eta$ for $6$~MeV $p$ impact on Li$(2s)$ for ejected electron energies $\eps$ of (a) 2~eV, (b) 10~eV, (c) 20~eV. The EXB1 and CP results are from Ref.~\cite{wal_14}, the experimental data are from Ref.~\cite{LaF_13}.}
\label{fig:lip_2s}
\end{figure}

\begin{figure}[ht]
    \centering
    \begin{subfigure}{0.48\textwidth}
        \includegraphics[width=\textwidth]{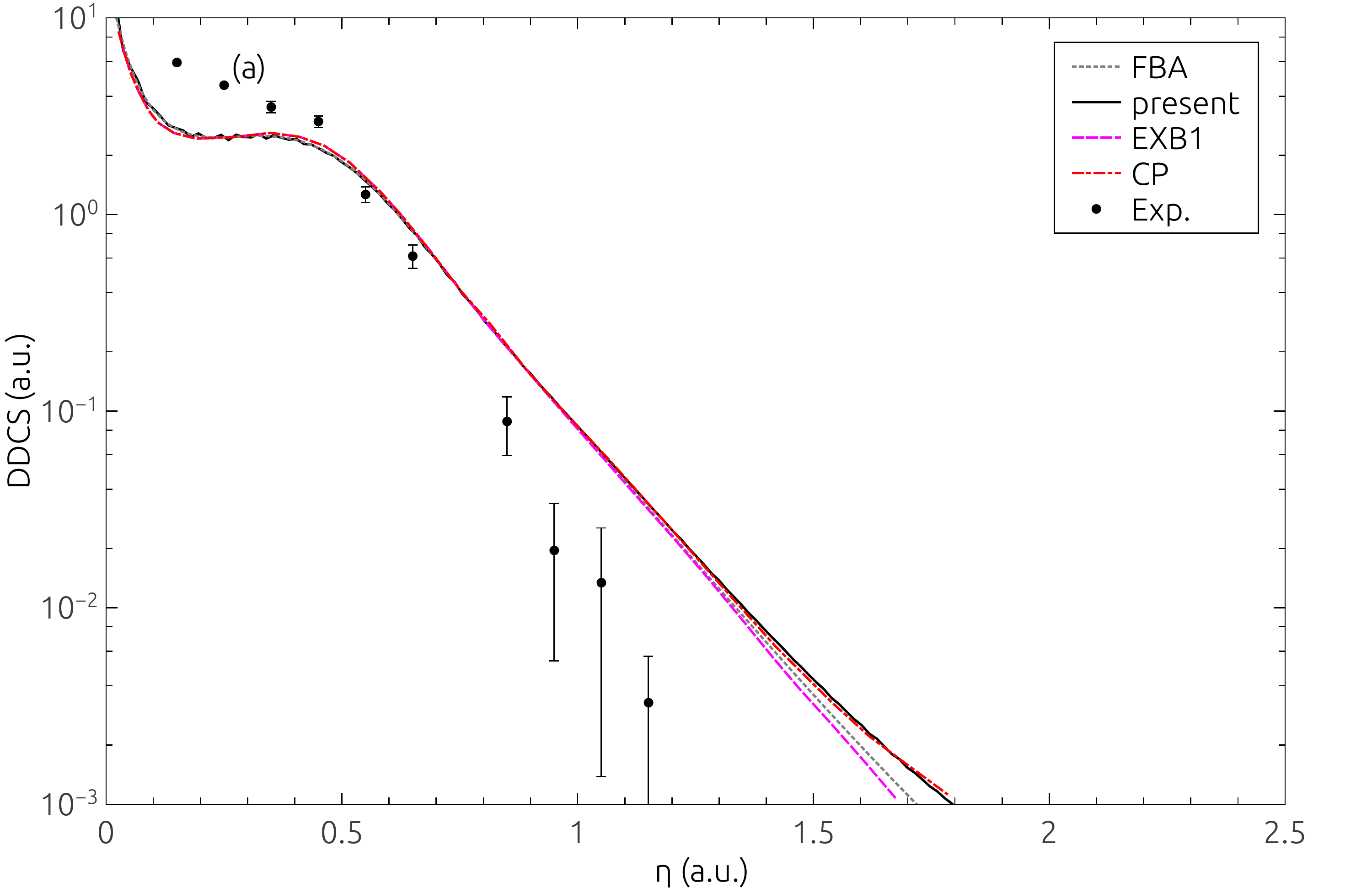}
    \end{subfigure}
\\ 
    \begin{subfigure}{0.48\textwidth}
        \includegraphics[width=\textwidth]{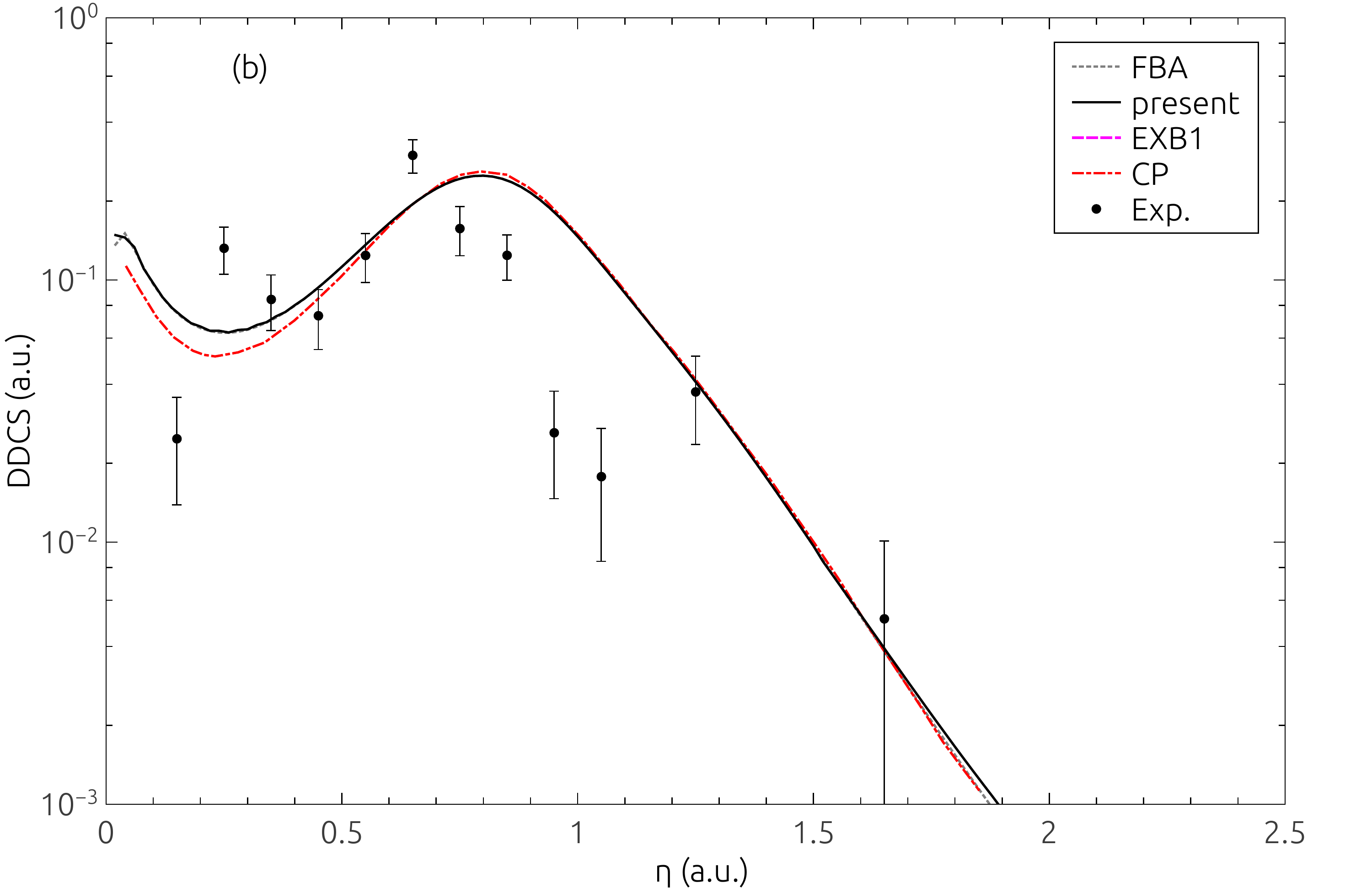}
    \end{subfigure}
\\
    \begin{subfigure}{0.48\textwidth}
        \includegraphics[width=\textwidth]{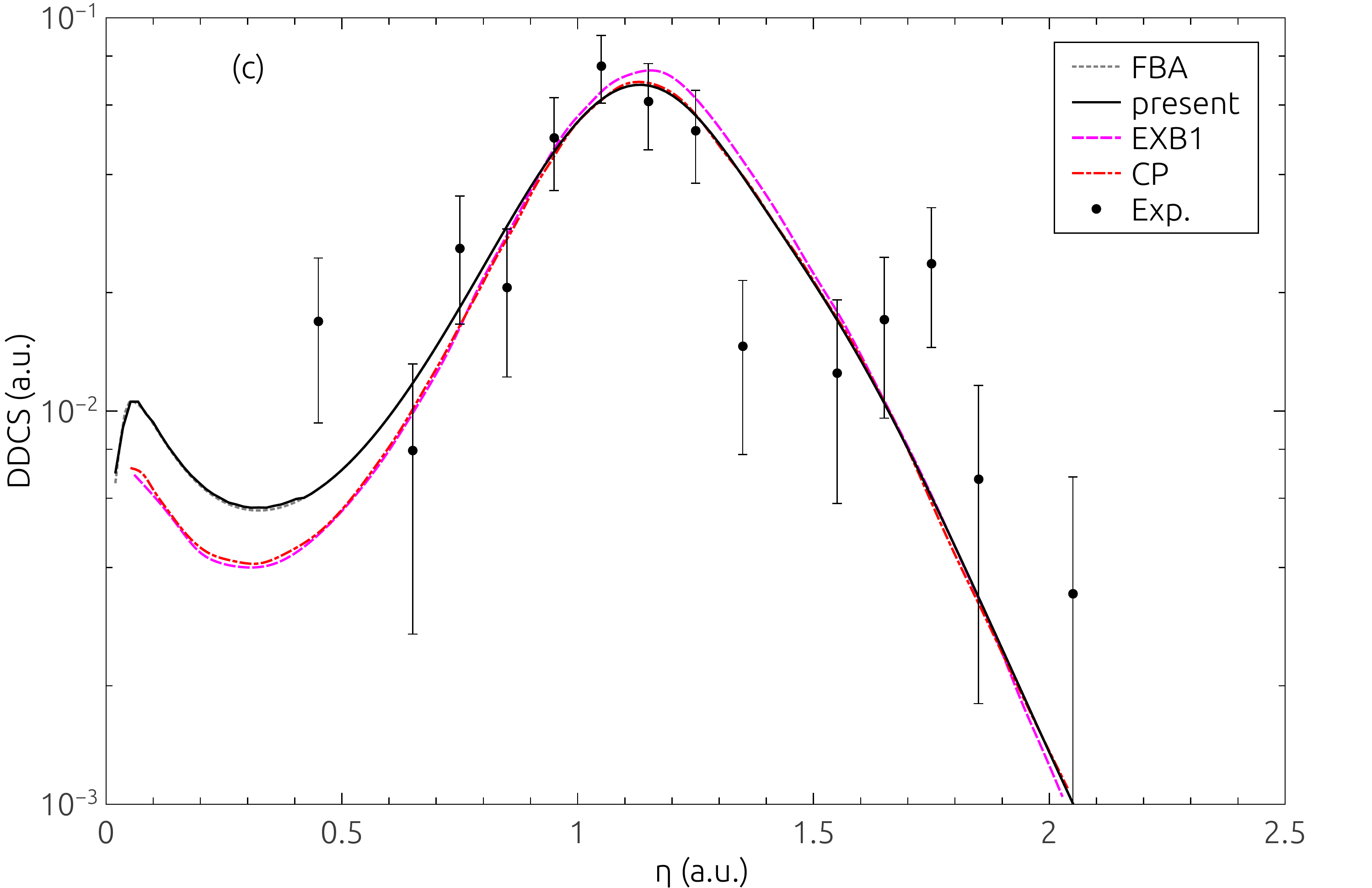}
    \end{subfigure}
    \caption{Same as Fig.~\ref{fig:lip_2s} but for $6$~MeV $p$ impact on Li$(2p)$.}
\label{fig:lip_2p}
\end{figure}

Figs.~\ref{fig:lip_2s} and~\ref{fig:lip_2p} show the DDCS $\frac{d^2\sigma}{d\eps\, d\eta}$ for the electron ejection energies of $2$, $10$, and $20$~eV in the $p$-Li($2s$) and $p$-Li($2p$) collisions at $6$~MeV, respectively. The EXB1 and CP results of Walters and Whelan~\cite{wal_14} as well as the experimental data of Laforge {\it et al.}~\cite{LaF_13} are shown for comparison. 
The results of our coupled-channel calculations in the first Born mode agree with those obtained in the wave treatment FBA 
provided that the equal number of partial waves is taken into account in both calculations. In turn, the convergence in the wave treatment FBA is well under control, since the ionization amplitude~\eqref{eq:pw_exp} is easy to calculate with a demanded precision. The difference between the coupled-channel calculation and FBA result at large $\eta$ is solely due to the NN interaction, which contributes to the former only. The results of the coupled-channel calculation neglecting the NN interaction [$\delta(b) \equiv 0$ in Eq.~\eqref{eq:FT_gen}] do not differ from the outcome of the FBA and are not shown in the figures. One can see that our results are in good agreement with the CP ones except for the region of small values of $\eta$. In this region, our results are systematically larger than the CP ones. This difference remains at the FBA level and, hence, can be attributed to the using of the different screening potentials for the lithium core. However, the difference is small, and both approaches are equally good in describing the experimental data, which are almost totally in the first Born regime.
%
\subsection{O$^{8+}$-impact ionization}  \label{s:oxygen-imp}
\begin{figure}[htb]
    \includegraphics[width=\linewidth]{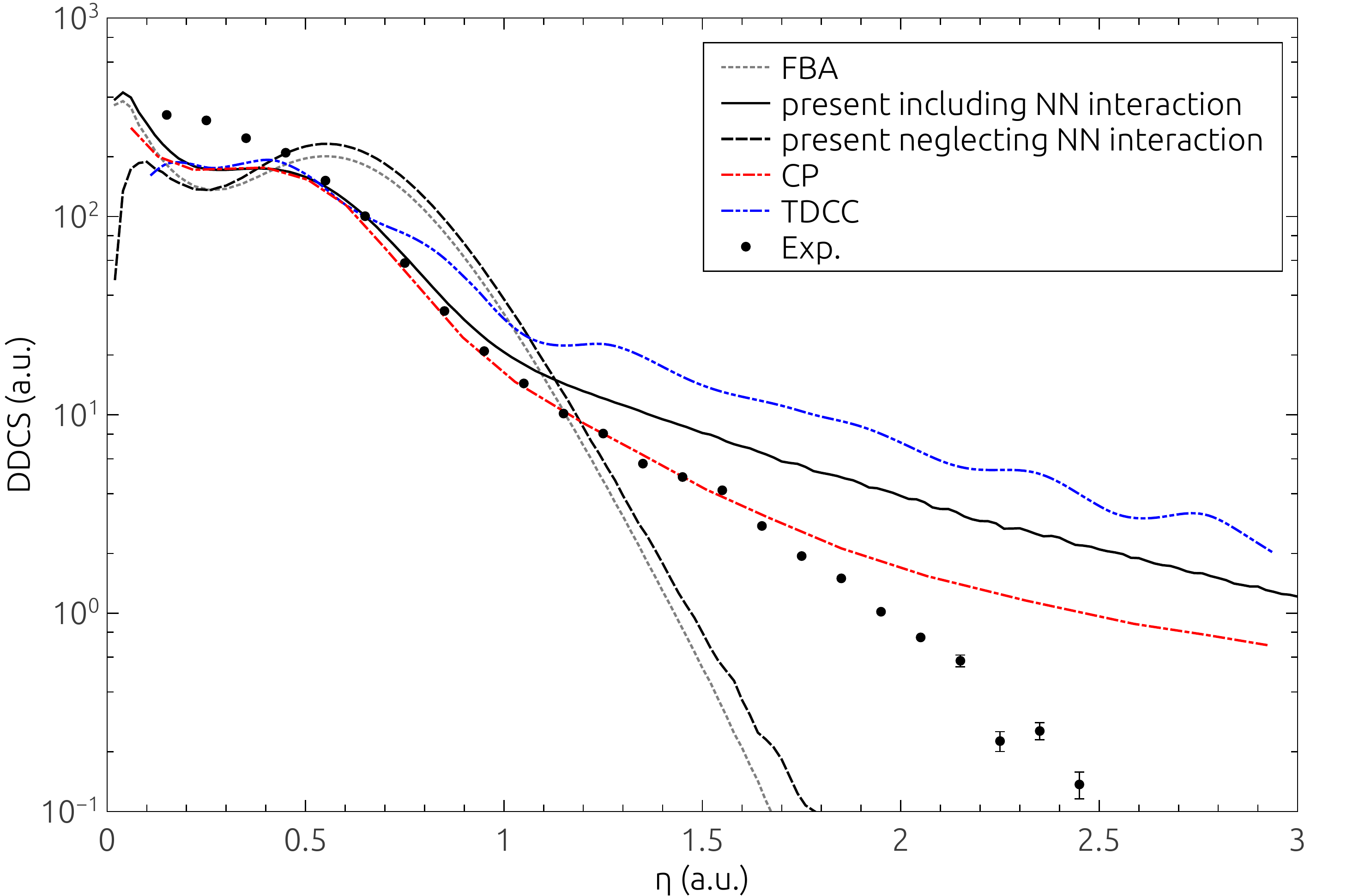}
    \caption{DDCS $\frac{d^2\sigma}{d\eps\, d\eta}$ as a function of the transverse component of the momentum transfer $\eta$ for $1.5$~MeV/u O$^{8+}$ impact on Li$(2s)$ for 2~eV electron ejection. The TDCC results are from Ref.~\cite{cia_13}, the CP results are from Ref.~\cite{wal_14}, and the experimental data are from Ref.~\cite{LaF_13}.}
\label{fig:liO_2s}
\end{figure}

\begin{figure}[htb]
    \includegraphics[width=\linewidth]{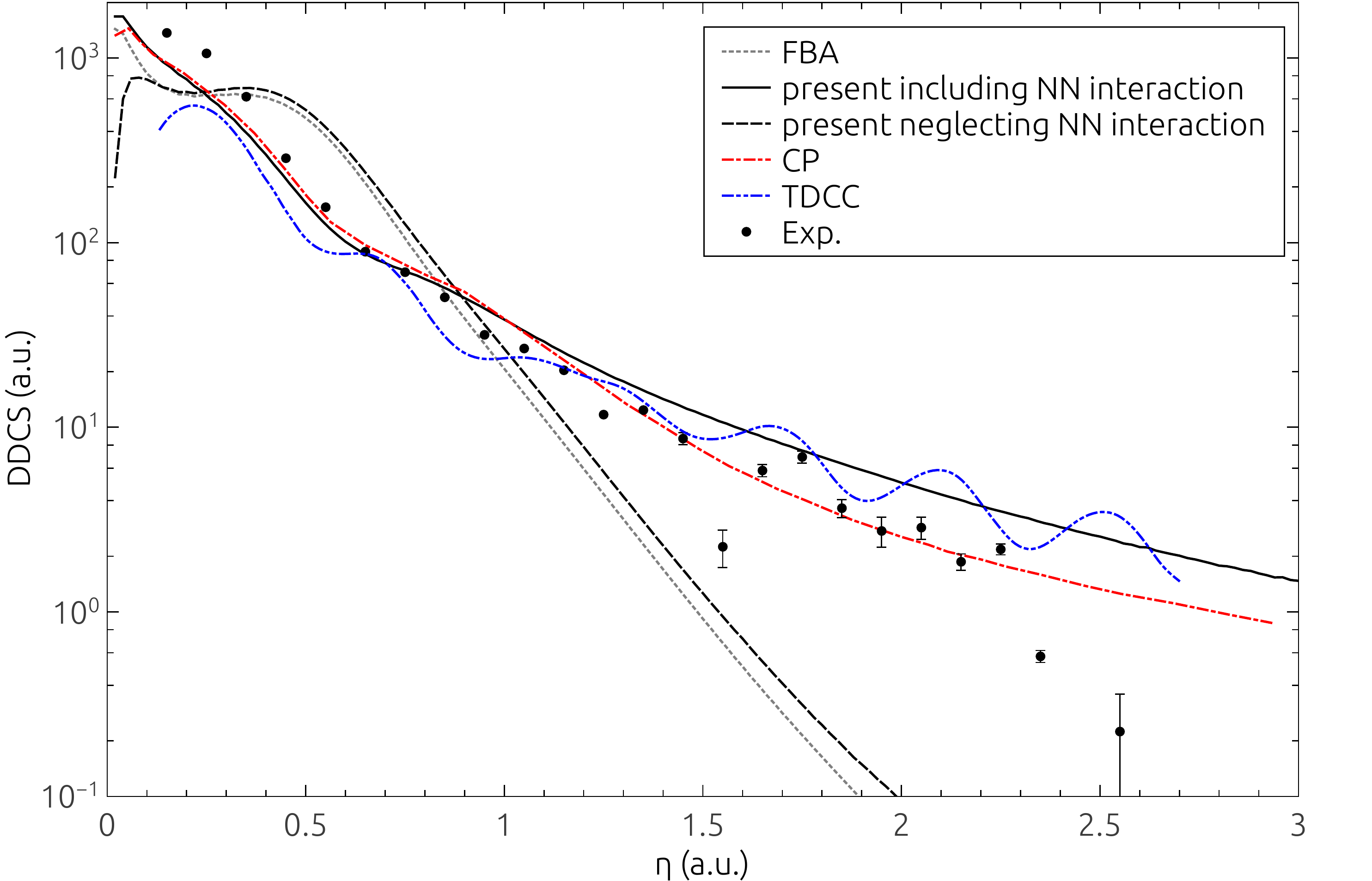}
    \caption{Same as Fig.~\ref{fig:liO_2s} but for $1.5$~MeV/u O$^{8+}$ impact on Li$(2p)$.}
\label{fig:liO_2p}
\end{figure}

%
In Figs.~\ref{fig:liO_2s} and~\ref{fig:liO_2p}, we present the DDCS $\frac{d^2\sigma}{d\eps\, d\eta}$ for the $2$-eV electron ejection in the O$^{8+}$-Li($2s$) and O$^{8+}$-Li($2p$) collisions at $1.5$~MeV/u, respectively. Our results are compared with the TDCC results~\cite{cia_13}, CP results~\cite{wal_14}, and experimental data~\cite{LaF_13}.
In the differential cross sections of Ref.~\cite{cia_13}, an extra factor is missed~\cite{pin_pc}, which leads to the normalization issue. Hence, we normalized the DDCS of Ref.~\cite{cia_13} to our results at $\eta = 0.65$~a.u. However, in this case, the normalization factors for the  DDCS from the $2s$ and $2p$ states were different, in contrast to the normalization of the experimental data, where the same factor was used.
From the figures, one can observe that the inclusion of the NN interaction is crucial for describing the experimental data. Neglecting the NN interaction in the calculation leads to the almost FBA behavior at large momentum transfers, which clearly contradicts with the measurements. The results of Walters and Whelan~\cite{wal_14} are closer to the experimental data than ours, which can be due to a more sophisticated form of the NN interaction used in their CP calculations. The normalized TDCC results of Ciappina {\it et al.}~\cite{cia_13} lie higher than ours for the ionization from the $2s$ state and oscillate about ours for the ionization from the $2p$ state. Similar oscillating behavior has been predicted by the TDCC approach for the DDCS in the antiproton-hydrogen collision~\cite{cia_13b}.
\section{Conclusions}  \label{s:conclusion}
In this paper, we have calculated the doubly differential cross sections for single ionization of lithium atom by $6$~MeV protons and $1.5$~MeV bare oxygen nuclei within the recently developed non-perturbative approach based on the Dirac equation~\cite{bon_17}. The results of the calculations are compared with the predictions of other non-perturbative approaches, such as the TDCC~\cite{cia_13} and CP~\cite{wal_14}, and the experimental data~\cite{LaF_13}. For light targets like lithium, our relativistic approach obviously does not give any advantages compared to other non-perturbative approaches. Nevertheless, for the studied collisions, its predictions are close to the measurements and can be used for explaining the data. The NN interaction has to be necessarily included in the DDCS calculation for collisions with O$^{8+}$ projectiles.
In the further work, we will focus on the differential cross sections in collisions of heavy ions and atoms, where the relativistic effects are large. Such cross sections are in the study list of the facility for antiproton and ion research (FAIR)~\cite{fair,les_16}.
\begin{acknowledgement}
We thank M.~F.~Ciappina and M.~S.~Pindzola for valuable discussions. This work was supported by RFBR (Grants No. 17-52-53136, No. 18-03-01220, No. 18-32-00279, and No. 18-32-20063), SPbSU-DFG (Grants No. 11.65.41.2017 and No. STO 346/5-1) and the Ministry of Education and Science of the Russian Federation (Grant No. 3.1463.2017/4.6). A.I.B. acknowledges the support from the German-Russian Interdisciplinary Science Center (G-RISC) funded by the German Federal Foreign Office via the German Academic Exchange Service (DAAD) and the FAIR-Russia Research Center.
\end{acknowledgement}
\section*{Author contribution statement}
All the authors were involved in the preparation of the
manuscript. All the authors have read and approved the final manuscript.
\bibliography{biblio}{}
\bibliographystyle{unsrt}
\end{document}